\documentclass[lettersize,journal]{IEEEtran}
\usepackage{amsmath,amsfonts}
\usepackage{algorithmic}
\usepackage{array}
\usepackage[caption=false,font=normalsize,labelfont=sf,textfont=sf]{subfig}
\usepackage{textcomp}
\usepackage{stfloats}
\usepackage{url}
\usepackage{verbatim}
\usepackage{graphicx}
\usepackage{makecell}
\usepackage{booktabs} 
\usepackage{multirow} 
\usepackage[table]{xcolor} 
\usepackage{amsmath} 
\usepackage{multicol}

\usepackage[utf8]{inputenc}
\usepackage{booktabs}
\usepackage[table]{xcolor} 
\usepackage{array}

\def\BibTeX{{\rm B\kern-.05em{\sc i\kern-.025em b}\kern-.08em
    T\kern-.1667em\lower.7ex\hbox{E}\kern-.125emX}}
\usepackage{balance}
\begin{document}
\title{MA-DLE: Speech-based Automatic Depression Level Estimation via Memory Augmentation}

\author{{
Xuzhi Wang$^1$,~\IEEEmembership{Member,~IEEE,} \quad
Xinran Wu$^1$, \quad
Ziping Zhao$^{1}$ ,~\IEEEmembership{Member,~IEEE,} \quad
Jianhua Tao$^2$,~\IEEEmembership{Senior Member,~IEEE,} \quad 
Björn W. Schuller$^{3,4}$,~\IEEEmembership{Fellow,~IEEE} \quad \\
$^1$ Tianjin Normal University \quad
$^2$Tsinghua University \quad
$^3$Technical University of Munich \quad
$^4$Imperial College London \quad

\thanks{
			\noindent\hfill\rule{\dimexpr\textwidth/4\relax}{0.4pt}\hfill\null\par
			\vspace{1em}		
		\noindent Xuzhi Wang, Xinran Wu and Ziping Zhao are with the School of Computer and Information Engineering, Tianjin Normal University (TJNU), Tianjin 300387, China (e-mail: wangxuzhi@tjnu.edu.cn; wxr@stu.tjnu.edu.cn; ztianjin@126.com).		
		 
		\noindent Jianhua Tao is with the Department of Automation, Tsinghua University, Beijing 100084, China (e-mail: jhtao@tsinghua.edu.cn). 
		
		\noindent Björn W. Schuller is the Chair of Health Informatics at the Technical University of Munich, Munich, Germany, and a Professor of Artificial Intelligence with the Department of Computing at Imperial College London, U.K. (e-mail: bjoern.schuller@imperial.ac.uk
).
		
		\noindent Corresponding author: Ziping Zhao.

        \noindent This work was supported by the National Natural Science Foundation of China under Grants No. 62071330, 61831022, U21B2020, and 62471249, the Humanities and Social Science Foundation of China Ministry of Education under Grant No. 24YJC740076, and the DFG (German Research Foundation) Reinhart Koselleck-Project AUDI0NOMOUS (Grant No. 442218748).
		
		}

}
}

\markboth{Journal of \LaTeX\ Class Files,~Vol.~, No.~, September~2025}%
{How to Use the IEEEtran \LaTeX \ Templates}

\maketitle

\begin{abstract}

Speech-based automatic estimation of depression levels is essential for enabling early detection and timely intervention, particularly in resource-constrained mental health settings. In recent years, deep learning has demonstrated impressive success across various domains, including affective computing and mental health assessment. Most existing approaches rely on RNN-based architectures (such as LSTM and GRU) to model temporal information for depression estimation. However, the extracted features often emphasize only a few adjacent speech segments, limiting their ability to capture long-range dependencies. To overcome this limitation, we introduce a memory-based feature augmentation method that enhances the representational capacity of GRU-extracted features. Rather than indiscriminately incorporating historical data, our memory bank is designed to selectively integrate two types of components in order to reduce redundancy and irrelevance: (1) historical temporal features that closely resemble the current GRU output, offering complementary contextual information; and (2) dynamic memory features identified based on feature variability, which capture behavioral and emotional fluctuations indicative of depressive symptoms. To effectively fuse the memory-augmented features with GRU outputs, we further design a Hierarchical Attention Fusion (HAF) module.  Our method is evaluated on the widely used DAIC-WOZ and E-DAIC datasets, achieving state-of-the-art performance.

\end{abstract}


\section{Introduction}
Depression is a common mental disorder that can cause individuals to experience persistent low mood, making it difficult for them to engage in daily social activities. In severe cases, it may even lead to suicide. Data released by the World Health Organization in 2017 indicates that approximately 350 million people worldwide suffer from depression. Furthermore, depression is projected to become the second leading cause of death by 2030~\cite{sharma2023depcap}. Traditional depression detection mainly relies on health questionnaires, which heavily depend on the subjective judgment of psychologists. This approach is not only time-consuming but also has limited accuracy, resulting in many patients failing to receive timely detection and treatment in the early stages. Moreover, especially in remote and economically underdeveloped areas, there is a severe shortage of psychological professionals, and the insufficient number of specialists makes it difficult for many individuals with depressive symptoms to obtain timely and professional diagnosis and treatment. Therefore, it is particularly urgent to develop automated depression monitoring systems to assist doctors in diagnosis. Such automated systems can efficiently process large volumes of data and conduct preliminary screening of large populations in a short period, significantly improving detection efficiency and coverage.

Many researchers have applied deep learning to the field of depression detection, primarily using GRU, LSTM and CNN models to capture temporal variations in speech. However, these approaches have certain limitations in modeling speech sequences, as they struggle to effectively capture long-range dependencies across time steps. As shown in the Fig.~\ref{fig: Temporal Similarity}, we compute the cosine similarity between speech signals from different time segments and the final output of the GRU. The results indicate that the final output focuses mainly on a few adjacent speech segments, lacking the ability to model speech information over longer temporal spans. This may lead to incomplete extraction of critical speech features, thereby affecting the accuracy of depression detection.

Inspired by the above findings, we propose a memory-based approach to capture long-term dependencies across speech segments. Such long-range dependencies are crucial for accurate depression level estimation for the following reasons: 1) The speech patterns of individuals with depression often exhibit long term dependencies. For example, variations in speaking rate, intonation, and pauses may evolve gradually over extended periods. 2) Depression is typically characterized by a sustained low emotional state, which cannot be adequately captured by a single short speech segment. 3) Short speech segments are susceptible to interference from environmental noise or momentary emotional fluctuations.

One major challenge in applying memory mechanisms to depression level prediction lies in the fact that speech signals contain a large amount of information unrelated to depression. If such irrelevant signals are incorporated without proper filtering, they may contaminate subsequent predictions. Although the output features extracted by GRU may not be sufficiently expressive, they still carry certain discriminative cues for depression. Therefore, on the one hand, it is beneficial to focus on historical features that are highly correlated with the current frame to provide supplementary information; on the other hand, even frames with relatively low correlation may contain critical discriminative signals and should not be completely discarded.

With these in mind, we introduce a Memory-Augmented Automatic Depression Level Estimation method, aiming to enhance the representation capability of GRU features, thereby promoting the task. Unlike the internal memory units in GRU, we propose an external structure with independent parameters to store various long-term features in the data that are beneficial for depression level estimation. Specifically, we first select features that are highly similar to the GRU output based on cosine similarity, treating them as semantic complements. Then, from the relatively dissimilar features, we extract temporal variation patterns to identify potentially informative cues for depression detection. Finally, We design a Hierarchical Attention Fusion (HAF) module to effectively leverage the complementary information embedded in the GRU outputs, similarity-retrieved features, and dynamic features.

The main contributions of our paper are as follows:

We propose a novel framework for  speech-based depression level estimation. To the best of our knowledge, this is the first work to introduce the memory bank mechanism into this task.

We propose a similarity-based feature retrieval approach to condense the memory bank and enhance it with dynamic features. These features are specifically designed to capture depressive cues, thereby improving the model’s ability to understand and predict depression levels.

We design a  Hierarchical Attention Fusion (HAF) module to effectively integrate the features from the memory bank and the GRU.

Our method achieves the state-of-the-art performance on DAIC-WOZ and E-DAIC datasets.

\section{Related Works}
  
    In recent years, as mental health issues have become increasingly prevalent, the early automatic detection of depression has emerged as a key research focus in multimodal affective computing. Emotional cues embedded in modalities such as speech, facial expressions, and textual language offer new opportunities for objective assessment. The following sections provide an overview of related work, primarily focusing on depression level estimation, while also covering aspects of depression detection. We discuss both traditional approaches based on handcrafted feature extraction and recent advancements in deep learning-based methods~\cite{sun2024novel, zhang2024multimodal, he2022deep, das2024deep, xu2024attention, xue2024fusing, fan2024transformer, xia2024depression}.

    \subsection{Traditional Approaches Based on Handcrafted Features}
        
        Handcrafted feature extraction has played an essential role in early research on automatic depression detection. These features are designed based on domain knowledge to capture relevant cues from different modalities. This section reviews representative studies based on handcrafted features from both speech and other modalities.

        In speech analysis, handcrafted features are designed to capture acoustic and prosodic variations that correlate with depressive symptoms. Prior work~\cite{jiang2018detecting} have explored the application of five types of handcrafted audio features in depression detection, including spectral features, cepstral features, glottal features, prosodic features, and voice quality features. These features can describe the low-frequency variations, intonation, speech rate, rhythm, and quality of speech, providing support for automatic depression detection.  Shin et al.~\cite{shin2021detection} employed a manual feature extraction method to extract four types of features from speech signals for depression detection, including glottal features, time-frequency features, formant features, and other physical features. These features are extracted individually within each speech segment and then averaged across the segment for subsequent analysis.

        Visual cues, particularly facial expressions and movements, also provide valuable information for depression detection. Previous works have extracted dynamic facial features using methods such as Local Phase Quantization (LPQ)~\cite{valstar2013avec,valstar2014avec}, Local Binary Pattern Three Orthogonal Planes (LBP-TOP)~\cite{dhall2015temporally}, Median Robust Local Binary Pattern (MRLBP)~\cite{he2018automatic}, and sparse coding~\cite{wen2015automated} to capture subtle non-verbal indicators of depression.

        Although handcrafted features have proven effective for depression detection and severity estimation, they heavily depend on expert design and may overlook subtle depression cues. Moreover, they often lack robustness across diverse individuals and fail to capture the temporal dynamics that are crucial for accurate depression assessment.

    \subsection{Data-Driven Methods with Deep Neural Networks}

    In recent years, with the rapid development of deep learning, an increasing number of studies have focused on depression prediction using architectures such as Convolutional Neural Networks (CNNs), Recurrent Neural Networks (RNNs), and Transformers. These approaches enable automatic feature learning from raw multimodal data, such as speech, text, and video, and have demonstrated promising performance in modeling both spatial and temporal patterns associated with depressive symptoms. These approaches offer a powerful and flexible alternative to traditional handcrafted feature-based techniques~\cite{yu2025using,li2025conformal,niu2025depression,niu2025examining,goncc2025affect,zhao2025multimodal,li2025efficient,su2026investigating,li2025automated,wang2025nuc-net,FFNet,ssc_MonoMRN,AdaSFormer2026,wang2023phase,luitel2025investigating, wu2025multimodal}.

    Some prior works have explored depression prediction based on speech. Han et al.~\cite{han2023spatial} introduced STFN, which employs VQWTNet for feature mapping, stacked gated residual blocks for multi-scale information. Chen et al.~\cite{chen2025ttfnet} proposed TTFNet, which encodes log-Mel spectrograms and their derivatives into quaternions, extracts frequency and temporal features, fuses them through XConformer blocks, and balances training with GradNorm. Zhang et al.~\cite{zhang2021depa} propose DEPA, a self-supervised audio embedding for depression detection, extracted using an encoder-decoder network on both in-domain (DAIC, MDD) and out-of-domain (Switchboard, Alzheimer’s) datasets.

    In addition to single-modality approaches, many studies have focused on multimodal fusion strategies to capture complementary information from different data sources~\cite{niu2025depression,zhao2025multimodal,li2025automated}. Guramritpal et al.~\cite{saggu2022depressnet} introduced DepressNet, a multimodal framework employing a hierarchical attention mechanism for depression detection. Their approach fuses multiscale temporal features from audio, video, and text modalities, leveraging a Bidirectional LSTM network and attention mechanisms for effective feature fusion. Marriwala et al.~\cite{marriwala2023hybrid} developed a hybrid deep learning model for depression detection, integrating text and audio features. The model combines Text CNN, Audio CNN, and hybrid LSTM/Bi-LSTM architectures for robust feature extraction and classification. Zhang et al.~\cite{zhang2025depitcm} proposed DepITCM, which integrates audio-visual features using an ITCM encoder, fuses time-channel-space information, and employs multi-task learning.

	Different from existing works, our approach is the first to introduce the memory mechanism into depression level estimation, aiming to address the forgetting issue commonly observed in GRU/LSTM-based models. Specifically, we propose enhancing the memory features through a combination of similarity-based feature retrieval and dynamic feature augmentation.

    \subsection{Memory-Augmented Networks}

    Existing memory-augmented recurrent neural networks (RNNs) can be broadly categorized into two types. The first type consists of models based on internal states, such as LSTMs and GRUs, which utilize hidden states and gating mechanisms to retain short-term and partially long-term information during sequence modeling. To our knowledge, most of existing methods~\cite{han2023spatial,chen2025ttfnet,rodrigues2019multimodal,yuan2024depression,uddin2022deep} for depression level estimation or detection fall into this category. The second type comprises models enhanced with external memory structures~\cite{xiao2024infllm,lee2021video,videnovic2025distractor,pang2025context}, which read from and write to external memory units to expand the network’s memory capacity, thereby enabling more effective capture of long-range dependencies or storage of complex structured information. Our approach belongs to this second category. External memory mechanisms have been relatively underexplored in speech processing. Emformer~\cite{shi2021emformer} introduces an efficient external memory that compresses long-range historical context into an augmented memory bank, significantly reducing self-attention computation for streaming ASR. Chen et al.~\cite{sheng2023face} addresses the Face-Driven Zero-Shot Voice Conversion task by employing a memory-based face-voice alignment module, in which memory slots act as a bridge to align the two modalities, enabling the extraction of voice characteristics from face images. While Emformer~\cite{shi2021emformer} primarily focuses on computational efficiency, Chen et al.~\cite{sheng2023face} emphasizes aligning multimodal face and audio information. In contrast, our approach leverages memory to capture rich emotional and temporal dynamics, which goes beyond modality alignment or efficiency optimization.

    To the best of our knowledge, our method is the first attempt of external memory mechanisms in the affective computing. Motivated by the limitations of GRU-based models in retaining long-range dependencies and the importance of temporal speech dynamics for depression detection, we introduce a similarity-based feature retrieval mechanism and a dynamic-feature memory module. Furthermore, we design a hierarchical attention fusion strategy to effectively integrate different types of memory representations.

\section{Preliminary and Analysis}

\begin{figure}[htbp]
   
        \centering
        % 插入PDF文件中的图像
        \includegraphics[width=\columnwidth]{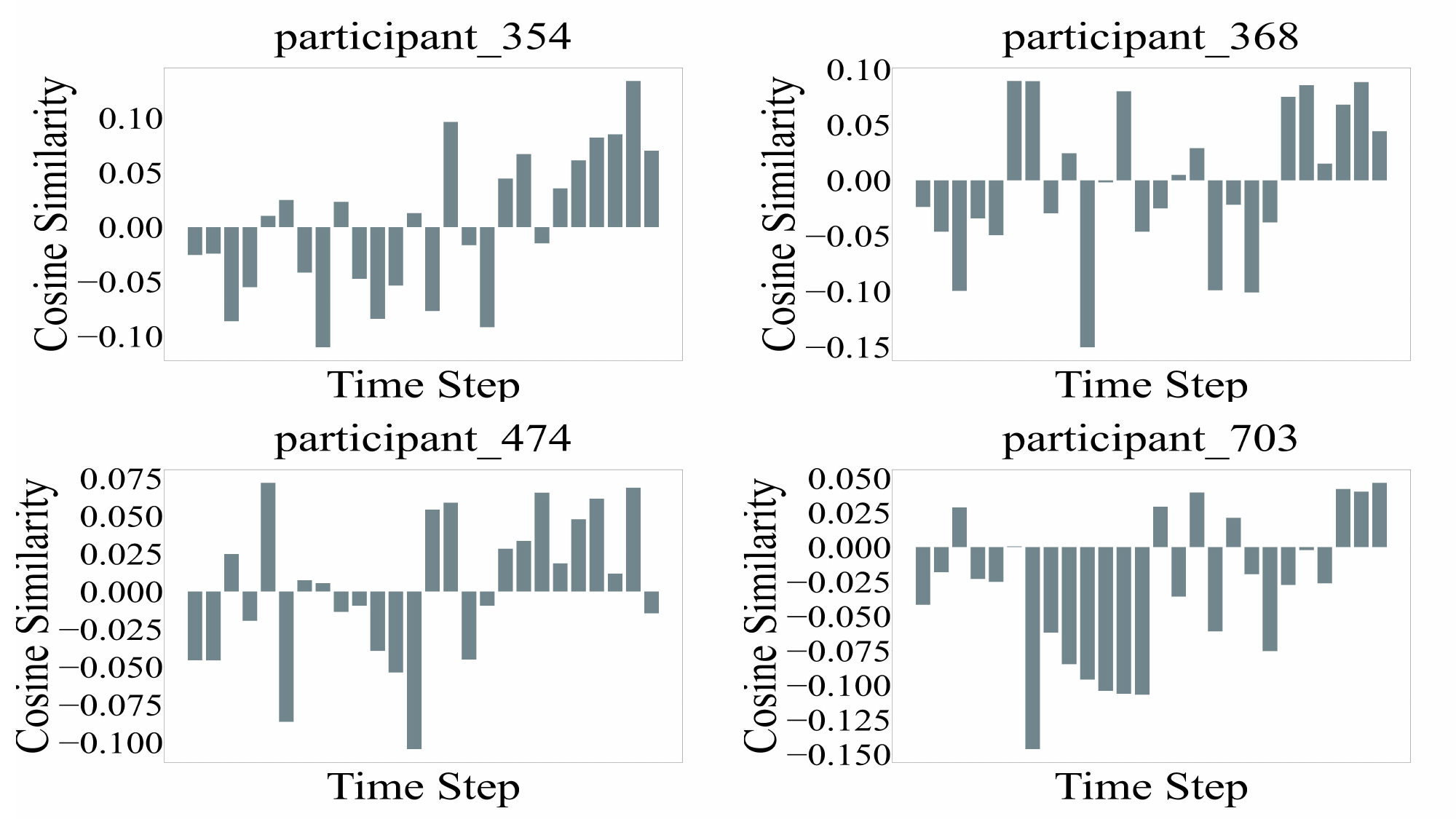} 
        \caption{The similarity between the GRU output and individual frame features. We can observe from these representative examples that the later frames tend to exhibit higher similarity with the GRU output. The x-axis represents the temporally downsampled time-dimension features, and the y-axis denotes the cosine similarity.} 
        \label{fig: Temporal Similarity}

\end{figure}

Recurrent Neural Networks (RNNs) are specifically designed to model temporal dependencies in sequential data. By maintaining a hidden state that captures information from previous time steps, RNNs are capable of learning patterns and relationships across time, making them well-suited for tasks involving time-series analysis, natural language processing, and other domains where context and order are crucial.

The Gated Recurrent Unit (GRU) is a variant of Recurrent Neural Networks (RNNs) that introduces gating mechanisms to better capture long-term dependencies and alleviate the vanishing gradient problem. Due to its efficiency and ability to model temporal patterns in sequential data, GRU has been widely applied in speech-based depression detection, where capturing subtle temporal cues in vocal signals is crucial for identifying depressive symptoms. The formulations of GRU are shown as follows:
\begin{align}
    \quad z_t &= \sigma(W_z \cdot [h_{t-1}, x_t]) ,\label{eq:update_gate} \\
    \quad r_t &= \sigma(W_r \cdot [h_{t-1}, x_t]) ,\label{eq:reset_gate} \\
     \quad \tilde{h}_t &= \tanh(W \cdot [r_t * h_{t-1}, x_t]), \label{eq:candidate_hidden} \\
     \quad h_t &= (1 - z_t) * h_{t-1} + z_t * \tilde{h}_t .\label{eq:hidden_state}
\end{align}

In Equation (4), the current hidden state $h_t$ is computed as a weighted sum of the previous hidden state 
$h_{t-1}$ and the current candidate state $\hat{h}_t$, where the weights are determined by the update gate $z_t$. If 
$z_t$ approaches 1, the current input (and the candidate state generated from it) has a stronger influence; conversely, if $z_t$ is close to 0, the influence of the historical state becomes dominant. As GRU updates its state step by step, historical information is progressively overwritten or updated by new inputs, especially those from recent time steps. This leads the model to become more sensitive to recent features, effectively assigning them greater weight.

We present a visualization of the cosine similarity between features extracted from different temporal segments and the GRU output. As shown in Fig.~\ref{fig: Temporal Similarity}, features from early temporal segments generally exhibit lower similarity with the GRU output and are more likely to show negative correlations, whereas features from later temporal segments tend to have higher similarity and exhibit more positive correlations. This phenomenon suggests that during sequential modeling, the GRU has a relatively limited capacity to retain information from early inputs and instead relies more heavily on the contextual information contained in later segments to form its final representation. This observation indirectly confirms the presence of a `forgetting' mechanism in GRU when processing long sequences, and further highlights the temporal imbalance in sequence modeling — that is, different temporal segments contribute unequally to the final output. 

Notably, since the GRU output and the features of each input frame represent different levels of semantic abstraction, their similarity tends to be low. Nevertheless, a consistent pattern can still be observed.

\begin{figure*}[htbp] % 使用 figure* 环境横跨两栏
    \centering % 图片居中
    \includegraphics[width=15cm]{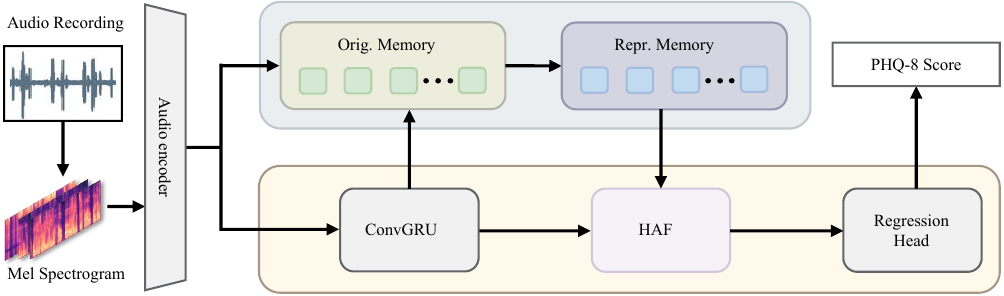} % 图片宽度设置为整个文本宽度
    \caption{The overall architecture of our method. The input audio signals are first transformed into Mel spectrograms, which are then processed by an audio encoder to extract high-level audio embeddings. These embeddings are subsequently fed in parallel into two branches. The upper branch incorporates a Memory Bank module, while the lower branch consists of a stack of ConvGRU modules. Then, the outputs of the two branches are fused through a Hierarchical Attention Fusion (HAF) module to obtain the final representation, which is passed through a regression head to predict the PHQ-8 score. `Orig. Memory' represents the original memory and `Repr. Memory' represents the representative memory. } % 图片标题
    \label{fig:framework} % 图片标签，用于引用
\end{figure*}

\section{Methods}

In this section, we first provide an overview of the problem formulation and our method. Next, we delve into the Memory Augmentation methods and transformer-based fusion mechanisms, which are the core of our method. Lastly, we introduce the loss function for model training.

\subsection{Overview}
\textbf{Problem Formulation.} Speech-based Depression level estimation can be formulated as follows. Let $\{A_i\}_{i=0}^{n}$ denote the input audio sequence consisting of frames from time step 0 to n. Let $y \in [0, 24]$ represent the PHQ-8 score, a clinically validated indicator of depression severity.  The goal is to learn a regression function $f$ that maps the input sequence to the target score, expressed as $y = f(\{A_i\}_{i=0}^{n})$.

\textbf{Overall Architecture.} We propose a novel method for depression level estimation based on memory augmentation. The overall framework of our approach is illustrated in Fig.~\ref{fig:framework}. Initially, audio signals are transformed into Mel spectrograms, which are then passed through NetVLAD to extract audio embeddings. The extracted speech feature embeddings are fed in parallel into two branches: the lower branch consists of a stack of ConvGRU modules, while the upper branch incorporates a Memory Bank module. The ConvGRU in the lower branch models the temporal variations of the speech signal, capturing sequential features that are closely associated with depressive symptoms. Meanwhile, the Memory Bank in the upper branch stores task-relevant similar features and dynamic features, serving as a crucial semantic supplement to the ConvGRU output and enhancing the model's ability to detect depression-related cues. Then, the GRU features, along with the task-relevant similar features and dynamic features, are fused using the proposed Transformer module. Finally, the fused representation is passed through a regression head to predict the PHQ-8 score.

\subsection{Memory Augmentation}

\begin{figure*}[htbp] % 使用 figure* 环境横跨两栏
    \centering % 图片居中
    \includegraphics[width=\textwidth]{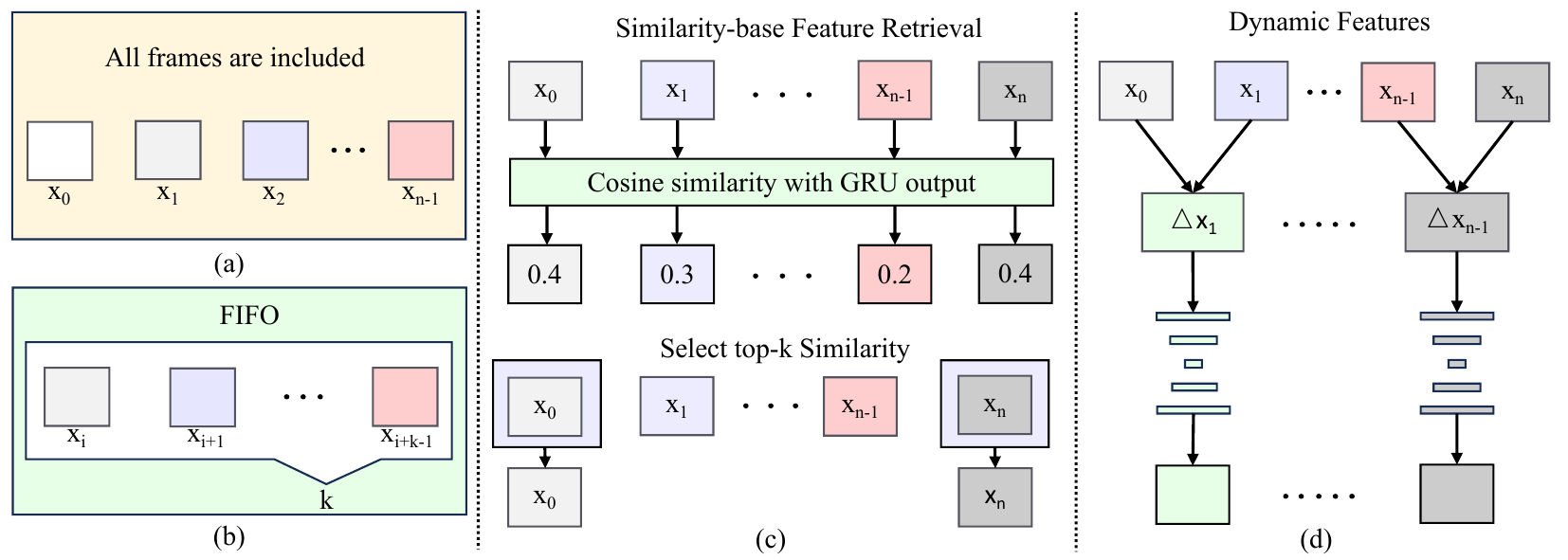} % 图片宽度设置为整个文本宽度
    \caption{Different memory bank construction methods.
(a) Speech features containing all time steps. (b) First-in-first-out (FIFO) strategy. (c) Augmenting with similarity-based feature retrieval, which first computes the similarity between each temporal feature segment and the GRU output, and then selects the top-k most similar features. (d) Augmenting with dynamic features, which capture temporal variations in speech that are indicative of depressive symptoms.} % 图片标题
    \label{fig:memory} % 图片标签，用于引用
\end{figure*}

In this paper, we introduce an explicitly designed external memory module—memory bank—into the task of depression level estimation, aiming to store long-term or global speech information. In preliminary experiments, we followed existing approaches by either writing all temporal speech features directly into the memory bank or updating its content dynamically using a first-in-first-out (FIFO) mechanism as shown in Fig.~\ref{fig:memory}. However, the former tends to introduce a large amount of redundant information, while the latter fails to retain long-term context, both of which limit the overall performance of the model.

To address the above issues, we filter the candidate features written into the memory bank based on their similarity to the GRU output features, effectively removing information irrelevant to depression assessment and retaining complementary representations. In addition, we extract the temporal variation patterns of speech features and incorporate them into the memory bank, thereby enhancing its ability to capture dynamic characteristics. The detailed method is described as follows.

\textbf{Augmenting with Similarity-Based Feature Retrieval.}
The features output by the GRU are closely related to depression level estimation. However, due to the GRU’s tendency to emphasize later frames in sequence modeling, important information in the earlier part of the sequence may be forgotten. To address this, we compute the similarity between early-frame features and the GRU output, and select those early features that exhibit high similarity. These selected features, which contain valuable information for depression level estimation, serve as an effective complement to the GRU output.

Formally, given the original speech feature sequence $X = \{x_1, x_2,...,x_T\}$ (the output of the audio encoder in Fig.~\ref{fig:framework}), where each $x_t \in \mathbb{R}^d$. We first calculate the cosine similarity between the output of GRU and all the temporal speech features as:
\begin{align}
    s_i = \text{sim}(\mathbf{q}, \mathbf{x}_i) = \frac{\mathbf{q}^\top \mathbf{x}_i}{\|\mathbf{q}\| \cdot \|\mathbf{x}_i\|}, \quad \text{for } i = 0, 2, \dots, T-1,
\end{align}
where $\mathbf{q}$ denotes the output feature of the GRU, and $s_i$ represents the similarity between $\mathbf{q}$ and the $i$-th feature in speech feature sequence.

Then, we select the top-K most similar features across time, which can be interpreted as the most relevant information for depression evaluation and a complementary representation to the GRU-extracted features:
\begin{align}
\mathcal{M}_K = \text{TopK}\left(\{s_i\}_{i=1}^N, K\right),
\end{align}
where $\mathcal{M}_K$ denotes the set of the top-K most similar features, and $\text{TopK}(\cdot, K)$ refers to the function that selects the K features with the highest similarity scores.

\textbf{Augmenting with Dynamic Features.} Besides incorporating features similar to the GRU outputs as complementary information, the temporal dynamics of speech features provide another important source of information. Temporal changes in speech features can reveal behavioral and emotional fluctuations, which are indicative of depressive symptoms.

Let $X = \{x_1, x_2,...,x_T\}$ denote the original speech feature sequence (the output of the audio encoder in Fig.~\ref{fig:framework}), where each $x_t \in \mathbb{R}^d$. To capture temporal variation, we perform frame-wise differencing to computing the changes in speech features, as shown below.
\begin{align}
    \Delta X = \{ x_t - x_{t-1} \mid t = 1, 2, \ldots, T-1 \},
\end{align}
where $\Delta X$ represents the sequence of frame-wise differences.

While frame-wise differencing provides a straightforward representation of first-order temporal changes between adjacent frames, it lacks the capacity to capture long-range dependencies and complex dynamic patterns, such as progressively increasing pitch or emotional shifts. To address this limitation, we design a lightweight temporal variation encoder to model the temporal dynamics of speech features.

In our experiments, we observe that directly feeding the entire difference sequence into the temporal variation encoder fails to yield satisfactory results. This is mainly because the dynamic variations in depressive speech are often extremely subtle and localized. When the difference features are encoded as a whole sequence, these transient cues tend to be smoothed out by adjacent frames, making them harder to detect. 

To address this issue, we propose a more fine-grained modeling approach that focuses on capturing local temporal dynamics.

(1) Frame-wise Modeling Strategy. To better preserve and model these local dynamics, we separate the difference features along the temporal dimension and feed each time-step individually into the temporal variation encoder. This frame-wise modeling strategy allows the encoder to focus on capturing fine-grained, localized temporal variations that are critical for depression detection. The formulation is shown below:
\begin{align}
    z_t = f_{dyn}(\Delta x_t), t=1,2,...,T-1,
\end{align}
where $f_{dyn}$ denotes the temporal variation encoder, which is designed to model higher-order dynamics from the frame-wise difference sequence $\Delta x_t$, and $z_t$ is the resulting latent representation that captures the temporal variation patterns of the speech features.

In addition to its general architectural layout, the Temporal Variation Encoder integrates two important component-level considerations:

(2) Max Pooling. Depressive speech often exhibits subtle dynamic variations that can be easily masked by environmental noise or speaker-specific characteristics. To mitigate this, we insert a max pooling layer after the convolutional blocks. This helps suppress low-amplitude or unstable fluctuations while preserving informative variations that may carry depressive cues.

(3) Batch Normalization. We remove batch normalization layers from the encoder. Due to the small batch sizes typically used in depression detection tasks and the high variability of individual speech patterns, batch normalization may introduce instability and hinder generalization. Eliminating it enables more stable and consistent feature extraction under diverse input conditions.

Then, we stack the frame-wise features to form the final dynamic representation $Z$.
\begin{align}
    Z = \mathrm{Concat}(z_1, z_2, \ldots, z_{T-1}) \in \mathbb{R}^{(T-1) \times D'},
\end{align}
where $D'$ represents the dimensionality of the output from the dynamic encoder and $\mathrm{Concat}(\cdot)$ denotes the concatenation of multiple vectors along the temporal dimension.

\begin{figure}[t!]   
        \centering
        \includegraphics[width=\columnwidth]{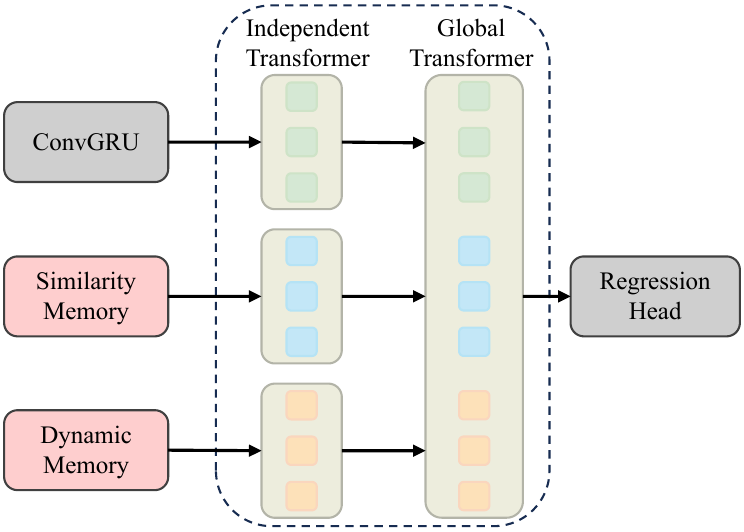} 
        \caption{Hierarchical Attention Fusion (HAF) module. The GRU output feature $q$, the similarity-retrieved feature $M_{K}$, and the dynamic feature $Z$ are first processed by independent Transformer blocks. Their resulting representations are then concatenated and fed into a global Transformer to enable hierarchical fusion of heterogeneous feature types.} 
        \label{fig:HAF}
\end{figure}
\subsection{Hierarchical Attention Fusion}
Due to the inherently different nature of the three types of features, including the GRU output, the similarity retrieved features, and the dynamic features, directly adding or concatenating them often leads to sub-optimal performance. Each type of feature captures distinct aspects of the input sequence: the GRU output encodes the global temporal context, the similarity-retrieved features emphasize historically relevant patterns, and the dynamic features highlight temporal variations indicative of depressive symptoms.

To address this issue, we propose a Hierarchical Attention Fusion (HAF) mechanism as shown in Fig.~\ref{fig:HAF}. Specifically, the GRU output feature $q$, the similarity-based retrieved feature $M_K$, and the dynamic feature $Z$
are first processed by three independent Transformer blocks, yielding the enhanced 
representations $q'$, $M_K'$, and $Z'$, respectively.
\begin{align}
    q' &= \mathcal{T}_q(q), \quad M_K' = \mathcal{T}_m(M_K), \quad Z' = \mathcal{T}_z(Z),
\end{align}
where $\mathcal{T}_q$, $\mathcal{T}_m$, and $\mathcal{T}_z$ denote three independent Transformer blocks.

Then, the processed features are concatenated and passed through another Transformer layer to enable global self-attention interaction and fusion. 
\begin{equation}
    H = \mathcal{T}_{\mathrm{global}}\Big( \mathrm{Concat}(q', M_K', Z') \Big)
\end{equation}
where $H$ denotes the globally fused output feature, and $Concat$ represents the concatenation operation along the feature dimension. This hierarchical design allows the model to fully exploit the complementary information embedded in each feature stream, leading to more discriminative representations for depression level estimation.

\subsection{Loss Function}
We introduce Smooth L1 Loss into depression level estimation. Smooth L1 Loss, also known as Huber Loss, behaves like Mean Squared Error (MSE) when the error is small, and like Mean Absolute Error (MAE) when the error is large. In the DAIC-WOZ and E-DAIC datasets, there exist a small number of extreme samples with PHQ-8 scores greater than 15, and in some cases, even exceeding 20. Introducing Smooth L1 Loss helps alleviate the instability in model training caused by these outlier samples. The formulation of the Smooth L1 Loss is shown as follows:
\begin{align}
    \text{SmoothL1}(x) &= 
    \begin{cases} 
    0.5 \cdot \frac{x^2}{\beta}, & \text{if } |x| < \beta \\
    |x| - 0.5 \cdot \beta, & \text{otherwise},
    \end{cases} 
\end{align}
where \( x \) represents the difference between the predicted value and the true value, and \( \beta \) is the smoothing parameter that determines the threshold between L2 (quadratic) and L1 (linear) behavior.

\section{Experiments}
\begin{table*}[htbp]
\caption{PHQ-8 Score Interpretation}
\centering
\begin{tabular}{ll>{\raggedright\arraybackslash}p{8cm}} 
\toprule 
\textbf{PHQ-8 Total Score} & \textbf{Depression Severity} & \textbf{Description} \\ \midrule 
0--4 & None or Minimal Depression & Generally considered within the normal emotional range \\ \addlinespace 
5--9 & Mild Depression & Possible depressive symptoms \\ \addlinespace
10--14 & Moderate Depression & Clinically significant; may benefit from counselling or intervention \\ \addlinespace
15--19 & Moderately Severe Depression & Professional help is advisable, such as therapy or medication \\ \addlinespace
20--24 & Severe Depression & Strongly suggests major depressive disorder \\ \bottomrule 
\end{tabular}
\label{tab:phq8}
\end{table*}
We conduct experiments on the widely used DAIC-WOZ and E-DAIC datasets, comparing our method with state-of-the-art approaches for depression level estimation. Subsequently, we provide a detailed analysis of the evaluation results, including comprehensive ablation studies. In addition, to further validate the effectiveness of the memory module, we perform t-SNE visualizations on the extracted features with and without memory.

\begin{table}[htbp]
    \centering
    \caption{Comparison with Previous Unimodal and Multimodal Methods on the DAIC-WOZ test set.}
    \label{tab:results2017}
    \begin{tabular}{lcccc} 
        \toprule
        \textbf{Method} & \textbf{Modality} & \textbf{Year} & \textbf{MAE} $\downarrow$ & \textbf{RMSE} $\downarrow$ \\
        \midrule
        \textcolor{gray}{Milintsevich et al.~\cite{milintsevich2023towards}} & \textcolor{gray}{T} & \textcolor{gray}{2023} & \textcolor{gray}{5.51} & \textcolor{gray}{-} \\
        \textcolor{gray}{Niu et al.~\cite{niu2021hcag}} & \textcolor{gray}{T} & \textcolor{gray}{2021} & \textcolor{gray}{3.73} & \textcolor{gray}{4.80} \\
        
        \addlinespace
        %\textcolor{gray}{Sun et al.~\cite{sun2017random}} & \textcolor{gray}{V} & \textcolor{gray}{2017} & \textcolor{gray}{4.89} & \textcolor{gray}{6.23} \\
        \textcolor{gray}{Rumahorbo et al.~\cite{rumahorbo2023exploring}} & \textcolor{gray}{V} & \textcolor{gray}{2023} & \textcolor{gray}{5.39} & \textcolor{gray}{6.27} \\
        \textcolor{gray}{Rathi et al.~\cite{rathi2019enhanced}} & \textcolor{gray}{V} & \textcolor{gray}{2019} & \textcolor{gray}{4.64} & \textcolor{gray}{5.98} \\
        
        \addlinespace
        \textcolor{gray}{Zhao et al.~\cite{zhao2019automatic}} & \textcolor{gray}{A+T} & \textcolor{gray}{2019} & \textcolor{gray}{4.20} & \textcolor{gray}{5.66} \\
        \textcolor{gray}{Lin et al.~\cite{lin2020towards}} & \textcolor{gray}{A+T} & \textcolor{gray}{2020} & \textcolor{gray}{3.75} & \textcolor{gray}{5.44} \\
        \addlinespace
        
        %\textcolor{gray}{Ringeval et al.~\cite{ringeval2017avec}} & \textcolor{gray}{A+V} & \textcolor{gray}{2017} & \textcolor{gray}{5.66} & \textcolor{gray}{7.05} \\
        \textcolor{gray}{Pan et al.~\cite{pan2023integrating}} & \textcolor{gray}{A+V} & \textcolor{gray}{2023} & \textcolor{gray}{4.62} & \textcolor{gray}{5.78} \\
        \textcolor{gray}{Li et al.~\cite{li2025audiovisual}} & \textcolor{gray}{A+V} & \textcolor{gray}{2025} & \textcolor{gray}{4.25} & \textcolor{gray}{5.34} \\
        \addlinespace
        
        \textcolor{gray}{Zhang et al.~\cite{zhang2022biomedical}} & \textcolor{gray}{A+V+T} & \textcolor{gray}{2022} & \textcolor{gray}{4.97} & \textcolor{gray}{6.45} \\
        \textcolor{gray}{Wei et al.~\cite{wei2022multi}} & \textcolor{gray}{A+V+T} & \textcolor{gray}{2022} & \textcolor{gray}{4.92} & \textcolor{gray}{5.86} \\
        \textcolor{gray}{Rasipuram et al.~\cite{rasipuram2022multimodal}} & \textcolor{gray}{A+V+T} & \textcolor{gray}{2022} & \textcolor{gray}{4.83} & \textcolor{gray}{5.76} \\
        \textcolor{gray}{Zhang et al.~\cite{zhang2024multimodal1}} & \textcolor{gray}{A+V+T} & \textcolor{gray}{2024} & \textcolor{gray}{4.48} & \textcolor{gray}{5.57} \\
        \textcolor{gray}{Fang et al.~\cite{fang2023multimodal}} & \textcolor{gray}{A+V+T} & \textcolor{gray}{2023} & \textcolor{gray}{-} & \textcolor{gray}{5.44} \\
        \addlinespace
        
        Jayawardena et al.~\cite{jayawardena2023ordinal} & A & 2023 & - & 6.84 \\
        Zhang et al.~\cite{zhang2021depa} & A & 2021 & 5.60 & 6.47 \\
        Han et al.~\cite{han2023spatial} & A & 2023 & 5.38 & 6.36 \\
        Zhang et al.~\cite{zhang2025depitcm} & A & 2025 & 5.21 & 6.10 \\
        %Williamson et al.~\cite{williamson2016detecting} & A & 2016 & 5.32 & 6.38 \\
        Qureshi et al.~~\cite{oureshi2021gender} & A & 2021 & 5.14 & 6.56 \\
        Alhanai et al.~\cite{al2018detecting} & A & 2018 & 5.13 & 6.50 \\
        Chen et al.~\cite{chen2025ttfnet} & A & 2025 & 5.09 & 6.01 \\
        Stepanov et al.~\cite{stepanov2018depression} & A & 2018 & 4.96 & 6.32 \\
        Yang et al.~\cite{yang2020feature} & A & 2020 & 4.63 & 5.52 \\
        Niu et al.~\cite{niu2025depression} & A & 2025 & 4.62 & 5.61 \\
        
        \addlinespace
        \hline
        Ours & A & 2025 & \textbf{4.31} & \textbf{5.49} \\ 
        \bottomrule
    \end{tabular}
\end{table}
 \begin{table}[htbp]
    \centering
    \caption{Comparison with Previous Unimodal and Multimodal Methods on the E-DAIC test set.}
    \label{tab:results2019}
    \begin{tabular}{lcccc}
        \toprule
        \textbf{Method} & \textbf{Modality} & \textbf{Year} & \textbf{MAE} $\downarrow$ & \textbf{RMSE} $\downarrow$ \\
        \midrule
        \addlinespace
        \textcolor{gray}{Zhang et al.~\cite{zhang2020multimodal}} & \textcolor{gray}{T} & \textcolor{gray}{2020} & \textcolor{gray}{-} & \textcolor{gray}{4.66} \\
        \textcolor{gray}{G{\"o}n{\c{c}} et al.~\cite{goncc2025affect}} & \textcolor{gray}{T} & \textcolor{gray}{2025} & \textcolor{gray}{3.46} & \textcolor{gray}{4.37} \\
        \addlinespace

         \textcolor{gray}{Xu et al.~\cite{xu2024two}} & \textcolor{gray}{V} & \textcolor{gray}{2024} & \textcolor{gray}{-} & \textcolor{gray}{5.99} \\
        \textcolor{gray}{Shen et al.~\cite{shen2024multi}} & \textcolor{gray}{V} & \textcolor{gray}{2024} & \textcolor{gray}{-} & \textcolor{gray}{5.83} \\
        \textcolor{gray}{Rodrigues et al.~\cite{rodrigues2019multimodal}} & \textcolor{gray}{V} & \textcolor{gray}{2019} & \textcolor{gray}{-} & \textcolor{gray}{5.74} \\
        \addlinespace
        
        \textcolor{gray}{Fan et al.~\cite{fan2019multi}} & \textcolor{gray}{A+T} & \textcolor{gray}{2019} & \textcolor{gray}{-} & \textcolor{gray}{5.91} \\
        \addlinespace  
        
        \textcolor{gray}{Ringeval et al.~\cite{ringeval2019avec}} & \textcolor{gray}{A+V} & \textcolor{gray}{2019} & \textcolor{gray}{-} & \textcolor{gray}{6.37} \\
        \textcolor{gray}{Fang et al.~\cite{fang2023multimodal}} & \textcolor{gray}{A+V} & \textcolor{gray}{2023} & \textcolor{gray}{-} & \textcolor{gray}{5.17} \\  
        \textcolor{gray}{Li et al.~\cite{li2025audiovisual}} & \textcolor{gray}{A+V} & \textcolor{gray}{2025} & \textcolor{gray}{4.41} & \textcolor{gray}{5.10} \\
        \textcolor{gray}{Pan et al.~\cite{pan2024disentangled}} & \textcolor{gray}{A+V} & \textcolor{gray}{2024} & \textcolor{gray}{4.32} & \textcolor{gray}{5.35} \\
     
        \addlinespace
        \textcolor{gray}{Yin et al.~\cite{yin2019multi}} & \textcolor{gray}{A+V+T} & \textcolor{gray}{2019} & \textcolor{gray}{-} & \textcolor{gray}{5.50} \\
        \textcolor{gray}{Saggu et al.~\cite{saggu2022depressnet}} & \textcolor{gray}{A+V+T} & \textcolor{gray}{2022} & \textcolor{gray}{-} & \textcolor{gray}{5.36} \\
        \textcolor{gray}{Zhang et al.~\cite{zhang2020multimodal}} & \textcolor{gray}{A+V+T} & \textcolor{gray}{2020} & \textcolor{gray}{-} & \textcolor{gray}{4.47} \\ 
        \textcolor{gray}{Sun et al.~\cite{sun2022cubemlp}} & \textcolor{gray}{A+V+T} & \textcolor{gray}{2022} & \textcolor{gray}{4.37} & \textcolor{gray}{-} \\       
        \textcolor{gray}{Yuan et al.~\cite{yuan2024depression}} & \textcolor{gray}{A+V+T} & \textcolor{gray}{2024} & \textcolor{gray}{3.98} & \textcolor{gray}{4.91} \\
        \addlinespace
        Sun et al.~\cite{sun2021multi} & A & 2021 & - & 8.67 \\
        Ringeval et al.~\cite{ringeval2019avec} & A & 2019 & - & 8.00 \\
        Rodrigues et al.~\cite{rodrigues2019multimodal} & A & 2019 & - & 6.71 \\
        Uddin et al.~\cite{uddin2022deep} & A & 2022 & - & 5.78 \\  
        Han et al.~\cite{han2023spatial} & A & 2023 & 5.38 & 6.29 \\
        Chen et al.~\cite{chen2025ttfnet} & A & 2025 & 5.00 & 5.76 \\
        %Yu et al.~\cite{yu2025using} & A & 2025 & - & 5.58 \\
        \addlinespace
        \hline
        Ours & A & 2025 & \textbf{4.68} & \textbf{5.72} \\ 
        \bottomrule
    \end{tabular}
\end{table}

 \subsection{Datasets}

\textbf{DAIC-WOZ Dateset.} The DAIC-WOZ~\cite{ringeval2019avec,ringeval2017avec,devault2014simsensei,gratch2014distress} dataset is a multimodal conversational dataset specifically designed for emotion and mental health analysis, primarily aimed at automatic depression detection and assessment. It was developed by the University of Southern California’s Institute for Creative Technologies to facilitate mental health evaluation through recordings of human-computer interviews, including audio, video, and textual data. The dataset contains a total of 189 interview sessions, including question-and-answer interactions from 59 depressed patients and 130 non-depressed patients. These samples are divided into three subsets: 107 sessions for training, 35 for validation, and 47 for testing. For technical resions, only 182 audio recordings are used. For depression level estimation, the dataset provides scores based on the Patient Health Questionnaire-8 (PHQ-8). PHQ-8 is a widely used clinical questionnaire consisting of eight items that measure depressive symptoms over the past two weeks. In this dataset, PHQ-8 scores are obtained from participants’ responses during structured interviews, serving as standardized labels for depression severity. The interpretation of the PHQ-8 scores is summarized in Table~\ref{tab:phq8}. PHQ-8 $\,\geq\,10$ is considered to indicate clinically significant depression and is commonly used as the threshold for determining the presence of depression.

\textbf{E-DAIC Dateset.} The E-DAIC dataset is an extended version of DAIC-WOZ, including 275 respondents. The training set includes 163 samples, the validation set includes 56 samples, and the test set includes 56 samples. The dataset includes data such as facial action units (Action Units, AU) and gaze coordinates (Gaze). Due to the multimodal data fusion characteristics of these two datasets and the fact that they contain a large amount of patient information in various forms, they are used in our experiment.

 \subsection{Evaluation Metrics}
In our experiment, we follow existing work and use Mean Absolute Error (MAE) and Root Mean Square Error (RMSE) as evaluation metrics for the regression task. MAE has high robustness, while RMSE amplifies the impact of larger errors through squaring operations, making it more sensitive to outliers. Smaller values of both metrics indicate that the prediction results are closer to the true values, and the model performs better. The formulations for MAE and RMSE are shown as follows:
 \begin{equation}
MAE = \frac{1}{T}\sum_{j = 1}^{T} \vert y_j - \bar{y}_j \vert \tag{5} ,
\end{equation}
\begin{equation}
RMSE = \sqrt{\frac{1}{T}\sum_{j = 1}^{T} (y_j - \bar{y}_j)^2} \tag{6} ,
\end{equation}
where T represents the total number of sample data, where \(y_j\) typically denotes the true value of the j-th sample and \(\bar{y}_j\) signifies its predicted value.

 \subsection{Experimental Setup and Training Configuration.}

Our implementation is based on PyTorch. We train the model for 500 epochs using the Adam optimizer with a learning rate of $1 \times 10^{-3}$ and weight decay of 1e-5, with the learning rate scheduled by CosineAnnealingLR and a batch size of 2. Temporal features are extracted using an 8-layer stacked GRU, and the number of Mel filter banks is set to 80 during preprocessing to capture perceptually relevant frequency components.

\subsection{Overview of the Model Architecture and Hyperparameter Settings.}
Table~\ref{tab:details of the proposed network} summarizes the key components of our model. The backbone ConvGRU module consists of an 8-layer unidirectional GRU with a hidden size of 256, an input dimension of 256, and a dropout rate of 0.7. The similarity-based memory module retrieves relevant historical features using cosine similarity followed by top-$K$ sampling, where $K=5$. The dynamic memory branch conducts frame-wise local modeling using a 1D convolution with a kernel size of 3 that expands the channel dimension from 1 to 12, followed by a ReLU activation and a max-pooling operation with a kernel size of 7. For the Independent Transformer, the hidden sizes are configured as follows: 256 for the GRU output feature $q$, 512 for the similarity-retrieved feature $M_K$, and 1024 for the dynamic feature $Z$. The Global Transformer block uses a hidden size of 512. After the Independent Transformer processing, the GRU output feature, as well as the similarity-retrieved and dynamic features, are further passed through a convolutional module to adjust their dimensionality.

\begin{table*}[htbp]
	\centering
	\caption{Details of the Key Components of Our Method.}
	\label{tab:details of the proposed network}
	\begin{tabular}{lcc}
		\toprule
		\textbf{Module} & \textbf{Operations} & \textbf{ Key Hyperparameters} \\
		\midrule
		
		ConvGRU & 8-Layer Unidirectional GRU & 
		\makecell[l]{
			Hidden Size = 256 \\
			Input Dimension = 256 \\
			Dropout Rate = 0.7
		} \\
		\hline
		
		Similarity-Based Memory & 
		\makecell[l]{Cosine Similarity \\ Top-K Sampling} & 
		K = 5 \\
		\hline
		
		Dynamic Memory & 
		\makecell[l]{
			Frame-Wise Modeling \\
			1D Conv + ReLU + Max Pooling \\
		} & 
		\makecell[l]{
			Conv Kernel Size = 3 \\
			Channels: 1 → 12 \\
			MaxPool Kernel Size = 7
		} \\
		\hline
		
		HAF & 
		\makecell[l]{
			Independent Transformer \\
			Global Transformer
		} & 
		\makecell[l]{
			Hidden Size = 256 / 512 / 1024 \\
			Hidden Size = 512
		} \\
		
		\bottomrule
	\end{tabular}
\end{table*}

 \subsection{Experimental Results}

To verify the effectiveness of the proposed method, we conduct experimental validation on two widely used depression detection datasets, DAIC-WOZ and E-DAIC. In the research, we compared the proposed method with existing other methods. The performance of our method and other state-of-the-art approaches on the DAIC-WOZ dataset is shown in Table ~\ref{tab:results2017}, and the results on the E-DAIC dataset are reported in Table~\ref{tab:results2019}.

It is important to note that our method is based solely on audio input. As many recent approaches rely on multimodal data, we include in our evaluation representative methods that utilize other modalities—including video-based, text-based, and multimodal approaches—to enable a more comprehensive and fair comparison.

As shown in Table~\ref{tab:results2017}, our method achieves an MAE of 4.31 and an RMSE of 5.49, outperforming all state-of-the-art audio-based methods. Moreover, it also yields competitive results compared to approaches using other input modalities. As shown in Table~\ref{tab:results2019}, which presents the performance on the E-DAIC dataset, our method achieves an MAE of 4.68 and an RMSE of 5.72, outperforming all audio-based state-of-the-art methods listed in the table and surpassing most methods based on other modalities

\subsection{Ablation Studies}

\begin{table}
    \centering
    \footnotesize
    \caption{Ablation study on the proposed memory bank on DAIC-WOZ test set.}

    \begin{tabular}{lcc}
    \toprule
    \textbf{Method} & \textbf{MAE} $\downarrow$ & \textbf{RMSE} $\downarrow$
    \\
    \midrule
    Baseline & 4.85 & 6.05        
    \\
    $+$ Full & 4.78 & 6.10        
    \\
    $+$ FIFO & 4.93 & 6.24        
    \\
     $+$ Sim & 4.56 & 5.89  
     \\
     $+$ Sim \& Dyn & \textbf{4.31}& \textbf{5.49}  
    \\

    \bottomrule
    \end{tabular}
    \label{tab:ablation_memory}
\end{table}

\begin{table}
    \centering
    \footnotesize
    \caption{Ablation study on the proposed memory bank on E-DAIC test set.}

    \begin{tabular}{lcc}
    \toprule
    \textbf{Method} & \textbf{MAE} $\downarrow$ & \textbf{RMSE} $\downarrow$
    \\
    \midrule
    Baseline & 5.16 & 6.32        
    \\
     w/o Sim & 4.89 & 5.99 
     \\
       w/o Dyn & 4.82 & 5.87 
     \\
       w/o HAF & 4.79 & 5.81 
     \\
     \hline
     Ours & \textbf{4.68}& \textbf{5.72} 
    \\

    \bottomrule
    \end{tabular}
    \label{tab:ablation_E-DAIC}
\end{table}
In this section, we conduct ablation studies to evaluate the effectiveness of the core components of our method. Specifically, we focus on the following aspects: (1) the impact of the proposed memory banks, (2) the design choices within the temporal variation encoder, (3) the design choices on loss function and (4) the design choices on fusion mechanisms.

\noindent\textbf{The Impact of the Proposed Memory Banks.} To investigate the contribution of the proposed memory banks, we conduct an ablation study by removing or modifying this component from our framework. The results on the DAIC-WOZ dataset are summarized in Table~\ref{tab:ablation_memory}, while the results on the E-DAIC dataset are presented in Table~\ref{tab:ablation_E-DAIC}. `Baseline' refers to a GRU-based model without incorporating any memory bank modules. `Full' represents the variant where the memory bank is constructed using features from all frames in the sequence, without temporal selection or compression. `FIFO' represents the variant where the memory bank is updated in a first-in-first-out manner, maintaining a fixed number of the most recent features. `Sim' denotes that We construct the memory bank by retrieving features from the sequence based on their similarity to the current query representation. `Dyn' represents the variant where inter-frame variation features are integrated into the memory bank. `AF' denotes the use of a Hierarchical Attention Fusion (HAF) mechanism to integrate the GRU output features, the similarity-based retrieved features, and the dynamic features.

As shown in Table~\ref{tab:ablation_memory}, on the DAIC-WOZ dataset, the baseline model that uses only GRU features achieves an MAE of $4.85$ and an RMSE of $6.05$. Including all frame features in the memory bank results in an MAE of $4.78$ and an RMSE of $6.10$, which does not lead to performance improvement. Using a FIFO strategy to update the memory bank results in an MAE of $4.93$ and an RMSE of $6.24$, leading to a decrease in performance. This may be due to the inclusion of redundant and irrelevant information. After retrieving features based on similarity before feeding them to the memory bank, the model achieves an MAE of $4.56\,(-0.29)$ and an RMSE of $5.89\,(-0.16)$, indicating a significant improvement. After incorporating dynamic features into the memory bank, the model achieves an MAE of $4.31\,(-0.54)$ and an RMSE of $5.49\,(-0.56)$.

As shown in Table~\ref{tab:ablation_E-DAIC}, on the E-DAIC dataset, the baseline achieves an MAE of $5.16$ and an RMSE of $6.21$. Without incorporating the similarity-based feature retrieval, the method yields an MAE of $4.89\,(+0.21)$ and an RMSE of $5.99\,(+0.27)$. Excluding the dynamic features from our method results in an MAE of $4.82\,(+0.14)$ and an RMSE of $5.87\,(+0.15)$. When the Hierarchical Attention Fusion is removed, the model obtains an MAE of $4.79\,(+0.11)$ and an RMSE of $5.96\,(+0.09)$. Overall, removing any of these three components leads to a performance drop, showing consistent effects on both the E-DAIC and DAIC-WOZ datasets.

\noindent\textbf{The Design Choices within the Temporal Variation Encoder.} To evaluate the impact of different design choices within the Temporal Variation Encoder, we perform a series of ablation studies. The experimental results are shown in Table~\ref{tab:ablation temporal variation encoder}. `Split' indicates that features from each frame are individually processed by the Temporal Variation Encoder. `Pooling' refers to the introducing the max pooling layers in the network to suppress low-amplitude or unstable fluctuations. `BN' represents the batch normalization layer.

As shown in Table~\ref{tab:ablation temporal variation encoder}, when each frame is not individually processed by the Temporal Variation Encoder, the model performance degrades, resulting in an MAE of $4.56 (+0.25)$  and an RMSE of $5.49 (+0.32)$. After removing the max pooling operation, the model performance declined, with the MAE increasing to $4.61(+0.30)$ and the RMSE to $5.67(+0.18)$. After introducing batch normalization, the model performance declined, with the MAE increasing to $4.49(+0.18)$ and the RMSE to $5.56(+0.07)$.

\begin{table}
    \centering
    \footnotesize
    \caption{The Design Choices within the Temporal Variation Encoder on DAIC-WOZ test set.}

    \begin{tabular}{lcc}
    \toprule
    \textbf{Method} & \textbf{MAE} $\downarrow$ & \textbf{RMSE} $\downarrow$
    \\
    \midrule
    With Split & \textbf{4.31} & \textbf{5.49}       
    \\
    W/o Split & 4.56 & 5.81  
    \\
    \hline
     With Pooling & \textbf{4.31} & \textbf{5.49}    
     \\
     W/o Polling & 4.61& 5.67 
     \\
     \hline
     With BN & 4.49& 5.56  
     \\
     W/o BN & \textbf{4.31} & \textbf{5.49}    
    \\

    \bottomrule
    \end{tabular}
    \label{tab:ablation temporal variation encoder}
\end{table}

\begin{table}
    \centering
    \footnotesize
    \caption{Ablation study on different fusion mechanisms on DAIC-WOZ test set.}

    \begin{tabular}{lcc}
    \toprule
    \textbf{Method} & \textbf{MAE} $\downarrow$ & \textbf{RMSE} $\downarrow$
    \\
    \midrule
    Addition & 4.57 &  5.74     
    \\
     Concatenation & 4.64 & 5.81
     \\
    Self-attention &4.69 &5.88 
    \\
    HAF &\textbf{4.31} &\textbf{5.49} 
    \\

    \bottomrule
    \end{tabular}
    \label{tab:ablation_fusion}
\end{table}

\noindent\textbf{The Design Choices on Fusion Mechanisms.} We conducted ablation studies to investigate the impact of different fusion mechanisms for integrating the GRU outputs with the memory-enhanced features. Our fusion strategy incorporates three types of features: the original GRU output, the similarity-based retrieved features, and the dynamic feature representations. `Addition' refers to the direct summation of the three types of features. `Concatenation' refers to the sequential concatenation of the three types of features. `Self-attention' refers to concatenating the three types of features first, followed by applying a Transformer-based self-attention mechanism over the combined feature representation. `HAF' refers to the proposed Hierarchical Attention Fusion mechanism, which first applies local self-attention to each of the three feature types individually, and then concatenates these features to perform a global self-attention over the combined representation. From Table~\ref{tab:ablation_fusion}, we observe that the HAF method achieves the best results. By employing a hierarchical design with staged fusion, it leads to more discriminative representations for depression level estimation, reducing the MAE by 0.26 and the RMSE by 0.25 compared to the best results of other methods.
\begin{figure*}[htbp]
    \centering
    % 插入PDF文件中的图像
    \includegraphics[width=15cm]{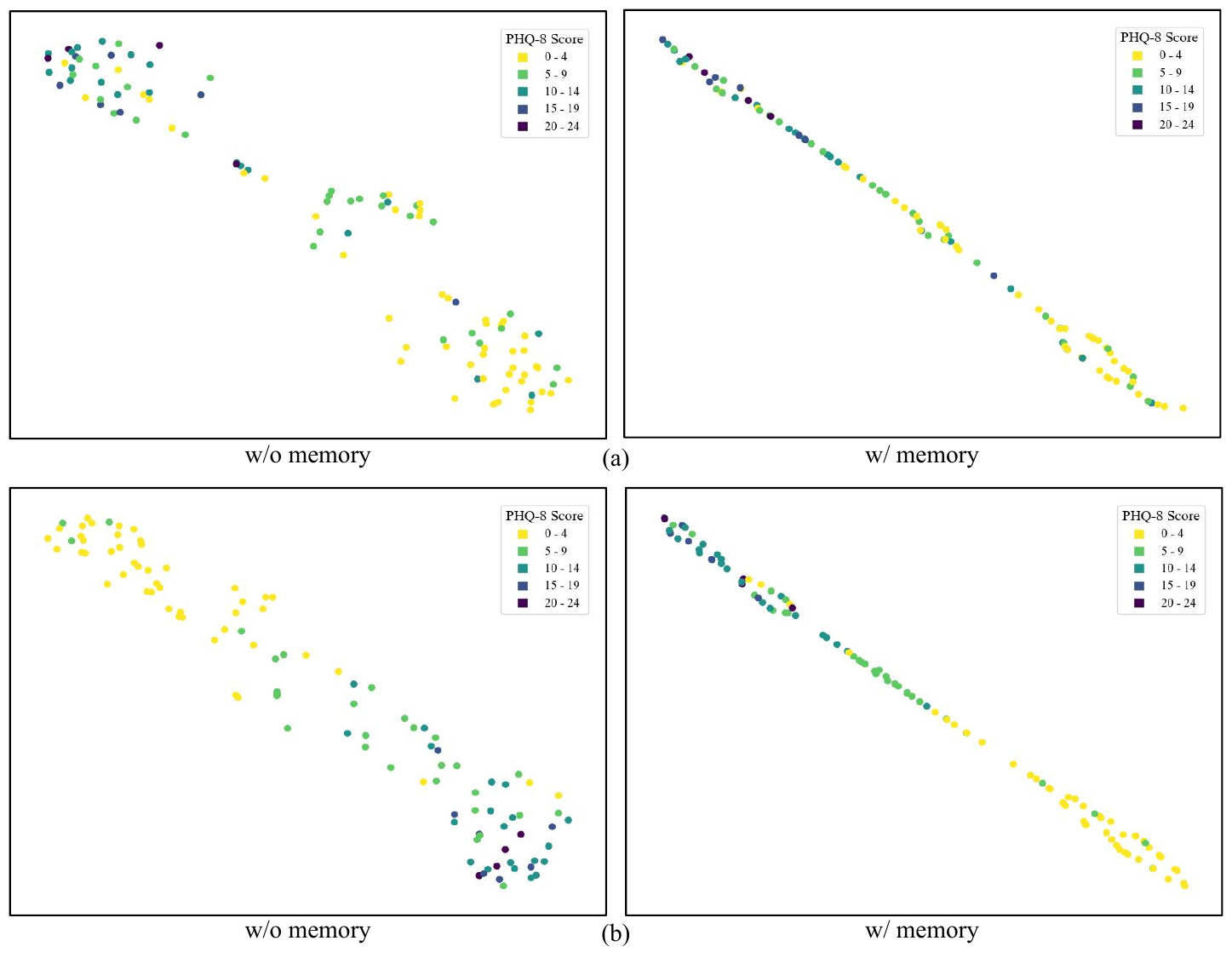} 
    \caption{The t-SNE visualization results of features with and without memory across different checkpoints. (a) shows the t-SNE visualization results at the middle stage of training, while (b) corresponds to the late stage. It can be observed that at both stages, the inclusion of memory features leads to better clustering of samples with similar levels of depression. This effect becomes more pronounced in the later stage of training, indicating that memory mechanisms help the model to learn more discriminative and compact feature representations over time.} % 标题放在左侧
    \label{fig:tSNE}
\end{figure*}

\noindent\textbf{The Design Choices on Loss Function.} To investigate the contribution of each component in the loss function to the overall model performance, we conducted an ablation study by selectively removing or modifying individual loss terms. In Table~\ref{tab:ablation_loss}, MAE indicates the use of mean absolute error loss, RMSE denotes the use of root mean square error loss, and Smooth L1 refers to the use of Smooth L1 loss. From Table~\ref{tab:ablation_loss}, we can observe that using the MAE loss results in a MAE of 4.39 and an RMSE of 5.84, while using the RMSE loss yields a MAE of 4.51 and an RMSE of 5.62. In comparison, the Smooth L1 loss achieves a MAE of 4.31 and an RMSE of 5.49, demonstrating improvements in both metrics.
\begin{table}
    \centering
    \footnotesize
    \caption{Ablation study on Loss Function on DAIC-WOZ test set.}

    \begin{tabular}{lcc}
    \toprule
    \textbf{Method} & \textbf{MAE} $\downarrow$ & \textbf{RMSE} $\downarrow$
    \\
    \midrule
    MAE & 4.39 &  5.84     
    \\
     RMSE & 4.51 & 5.62
     \\
     Smooth L1 &\textbf{4.31} &\textbf{5.49} 
    \\

    \bottomrule
    \end{tabular}
    \label{tab:ablation_loss}
\end{table}

\noindent\textbf{Ablation Analysis of Memory Mechanisms on Different Backbone Architectures.} We conduct extensive ablation studies across various backbone architectures, including GRUs with different depths, BiLSTMs, LSTMs, and Transformers. As shown in \ref{tab:memory mechanism across different backbone architectures}, the results consistently show that incorporating the memory-augmentation mechanism leads to improvements in model performance.

	\begin{table}[htbp]
	\centering
	\caption{Ablation studies of the memory mechanism across different backbone architectures on the DAIC-WOZ test set.}
	\label{tab:memory mechanism across different backbone architectures}
	\begin{tabular}{lcc}
		\toprule
		\textbf{Methods} & \textbf{MAE} $\downarrow$ & \textbf{RMSE} $\downarrow$ \\
		\midrule
		GRU 1 Layer w/o Memory & 4.86 & 6.37 \\
		GRU 1 Layer w/ Memory & 4.51 & 6.13 \\
		GRU 2 Layers w/o Memory & 4.91 & 6.28 \\
		GRU 2 Layers w/ Memory & 4.42 & 6.01 \\
		GRU 4 Layers w/o Memory & 4.89 & 6.13 \\
		GRU 4 Layers w/ Memory & 4.45 & 5.91 \\
		LSTM w/o Memory& 4.57 & 6.01 \\
        LSTM w/ Memory& 4.35 & 5.68 \\
        BiLSTM w/o Memory& 4.65 & 6.14 \\
        BiLSTM w/ Memory& 4.44 & 6.07 \\
        Transformer w/o Memory& 4.74 & 6.11 \\
        Transformer w/ Memory& 4.68 & 5.98 \\

		\addlinespace
		\hline
		GRU 8 Layers w/ Memory (Ours) & \textbf{4.31} & \textbf{5.49} \\
		\bottomrule
	\end{tabular}
\end{table}

\noindent\textbf{Sensitivity Analysis of the Top-K Parameter.} As shown in Table~\ref{tab:Top-K}, the experimental results show that the model achieves the best performance when $k=5$. When $k$ is too small, the retrieved information is insufficient to cover key depression-related cues. However, when $k$ is too large, it introduces excessive irrelevant or noisy information, which weakens the model's ability to focus on informative features.

	\begin{table}[htbp]
	\centering
	\caption{Sensitivity Analysis of the Top-K Parameter in Similarity Retrieval on the DAIC-WOZ test set.}
	\label{tab:Top-K}
	\begin{tabular}{lcc}
		\toprule
		\textbf{Methods} & \textbf{MAE} $\downarrow$ & \textbf{RMSE} $\downarrow$ \\
		\midrule
		
		K=1 & 4.56 & 5.99 \\
		K=3 & 4.41 & 5.68 \\
        K=5 (Ours) & \textbf{4.31} & \textbf{5.49} \\
        K=7 & 4.33 & 5.65 \\
        K=9 & 4.45 & 5.66 \\

		\bottomrule
	\end{tabular}
\end{table}

\noindent\textbf{Sensitivity Analysis of the $\beta$ Parameter in Smooth-L1 Loss.} As shown in Table~\ref{tab:Smotth-L1}, the experimental results indicate that the model achieves the best MAE performance when $\beta = 0.5$, while the best RMSE performance is obtained when $\beta = 1$. When $\beta$ is set to either a smaller or larger value beyond these ranges, both MAE and RMSE degrade modestly. This suggests that an inappropriate $\beta$ either over-emphasizes the L1 region (leading to insufficient sensitivity to moderate errors) or over-expands the L2 region (causing the model to over-penalize large errors), ultimately harming overall performance.

\begin{table}[htbp]
	\centering
	\caption{Sensitivity Analysis of the $\beta$ Parameter in Smooth-L1 Loss on the DAIC-WOZ test set.}
	\label{tab:Smotth-L1}
	\begin{tabular}{lcc}
		\toprule
		\textbf{Methods} & \textbf{MAE} $\downarrow$ & \textbf{RMSE} $\downarrow$ \\
		\midrule
		
		$\beta$=0.0 & 4.39 & 5.84 \\
        $\beta$=0.5 & \textbf{4.27} & 5.56 \\
		$\beta$=1.0 (Ours) & 4.31 & \textbf{5.49} \\
        $\beta$=1.5 & 4.42 & 5.67 \\
        $\beta$=2.0 & 4.68 & 5.98 \\

		\bottomrule
	\end{tabular}
\end{table}

\noindent\textbf{Impact of Historical Memory Length on Our Method}. In our ablation study, the GRU processes the full input sequence, while the memory module is evaluated using historical features of varying lengths. As shown in Table~\ref{tab:memory_length}, when the memory length increases from 5 to 25, the model performance steadily improves. Specifically, with a memory length of 5, the model achieves an MAE of 4.68 and an RMSE of 5.98, while increasing the memory length to 25 results in an MAE of 4.35 and an RMSE of 5.66. Once the memory length exceeds 25, the performance tends to stabilize, with only marginal gains observed.

        \begin{table}[htbp]
	\centering
	\caption{Sensitivity Analysis of the Historical Sequence Length on the DAIC-WOZ test set.}
	\label{tab:memory_length}
	\begin{tabular}{lcc}
		\toprule
		\textbf{Memory Length} & \textbf{MAE} $\downarrow$ & \textbf{RMSE} $\downarrow$ \\
		\midrule
		
		L=5 & 4.73 & 5.98 \\
		L=10 & 4.56 & 5.79 \\
        L=15  & 4.41 & 5.67 \\
        L=20 & 4.46 & 5.65 \\
        L=25 & 4.35 & 5.66 \\
        L=30 & 4.30 & 5.51 \\
        L=35 & 4.33 & \textbf{5.47} \\
        L=40 & \textbf{4.29} & 5.56 \\

		\bottomrule
	\end{tabular}
\end{table}

 \subsection{Model Complexity.} The proposed model contains 9.00M learnable parameters and requires only 0.72 GFLOPs for a single forward pass. Since few open-source methods are available, we compare only with Wei et al. [54], which uses 7.17M parameters but requires 7.18 GFLOPs per forward pass. The higher computational cost of Wei et al. is mainly attributed to its use of a larger number of sample slicing windows, which increases the overall processing overhead. This comparison highlights that our architecture achieves efficient inference with significantly lower computational cost while maintaining strong predictive performance.

 Moreover, our method exhibits excellent scalability to long input sequences, primarily because all core operations are designed to be computationally lightweight. For instance, each time-step audio feature and its corresponding GRU output both have a dimensionality of 256. Computing the cosine similarity between them requires only about 2k FLOPs, which constitutes a negligible computational cost in the overall pipeline.

\begin{table}[htbp]
\centering
\caption{Model complexity and performance comparison.}
\label{tab:example}
\begin{tabular}{lcccc}
\toprule
\textbf{Methods} & \textbf{Parameter}$\downarrow$ & \textbf{FLOPs}$\downarrow$ & \textbf{MAE}$\downarrow$ & \textbf{RMSE}$\downarrow$ \\
\midrule
%Shen et al.~\cite{shen2022automatic} & 0.921M & 0.0049G & 5.23 & 6.40 \\
Wei et al.~\cite{wei2022multi} & \textbf{7.17M} & 7.18G & 4.92 & 5.86 \\
Ours & 9.00M  & \textbf{0.72G} & \textbf{4.31} & \textbf{5.49} \\ 
\bottomrule
\end{tabular}
\end{table}

 \subsection{Visualizing the Impact of Memory Features.}

 In this section, we employ t-SNE to visualize the feature representations extracted by the model, aiming to compare the representational capacity of the model in the feature space with and without the incorporation of memory features. Since the depression severity scores are continuous in regression tasks and not directly suitable for categorical visualization, we categorize the samples into five levels based on standard depression severity criteria: non or minimal depression, mild depression, moderate depression, moderately severe depression, and severe depression. This categorization enables a more intuitive observation of the distribution of features across different severity levels.

As shown in Fig.~\ref{fig:tSNE}, the visualization results reveal that, without memory features, samples of different depression levels tend to overlap significantly in the feature space, resulting in blurred class boundaries. In contrast, after incorporating memory features, samples within the same category are more tightly clustered, and inter-class separation becomes more distinct. These observations indicate that the memory module enhances the model’s ability to capture temporal dependencies in the input sequences, thereby improving the quality of the learned feature representations and strengthening the model's capability to discriminate between different levels of depression severity.

\section{Conclusion}
In this work, we proposed a memory-enhanced framework to address the limitation of traditional GRU and LSTM models in retaining long-term temporal information, particularly the tendency to forget early-stage features. To this end, we introduced two key components: a similarity-based feature retrieval mechanism and a dynamic feature bank, designed to capture critical depression-related cues and enhance the model’s capability in estimating depression severity levels. Our method achieved state-of-the-art performance on both the DAIC-WOZ and E-DAIC datasets, demonstrating its effectiveness and generalizability. In future work, we plan to further explore the integration of multi-modal memory structures and investigate the applicability of our approach to broader mental health assessment tasks.

\bibliographystyle{IEEEtran}
\bibliography{IEEEabrv}

%\begin{IEEEbiographynophoto}
%\end{IEEEbiographynophoto}

\end{document}